\documentclass[10pt,letterpaper]{article}
\usepackage[T1]{fontenc}
\usepackage{lmodern,microtype}
\usepackage[margin=0.92in,top=0.79in,bottom=0.78in]{geometry}
\usepackage{amsmath,amssymb,booktabs,graphicx,tabularx,array}
\usepackage[font=small,labelfont=bf,labelsep=period]{caption}
\usepackage{xcolor}
\usepackage[colorlinks=true,allcolors=blue!45!black]{hyperref}
\usepackage{enumitem,placeins}
\newcommand{\doi}[1]{\href{https://doi.org/#1}{doi:\,#1}.}
\newcommand{\RR}{\mathbb R}
\newcommand{\wh}{\widehat}
\newcommand{\avg}{\operatorname{avg}}
\newcommand{\clip}{\operatorname{clip}}
\hypersetup{pdftitle={Which Vertebral Landmark Errors Change a Cobb Angle?},pdfauthor={Zizhao Peng},pdfkeywords={Cobb angle estimation, vertebral landmark regression, geometry-weighted loss, adolescent idiopathic scoliosis, spinal radiography}}
\title{\vspace{-1.8em}\Large\bfseries Which Vertebral Landmark Errors Change a Cobb Angle?\\[0.3ex]\large\mdseries Geometry-Weighted Supervision for Radiographic Landmark Regression}
\author{Zizhao Peng\\[0.35ex]\small Department of Computing, The Hong Kong Polytechnic University, Kowloon, Hong Kong\\\small \texttt{21051291r@connect.polyu.hk}}
\date{}
\begin{document}
\maketitle
\begin{abstract}
A network that localizes vertebral corners can reduce pixel error while rotating the line used to measure spinal curvature. We derive a training loss from the geometry of a three-angle reader. Of the eight corner-displacement degrees of freedom of a vertebral quadrilateral, one local direction rotates its left-to-right chord. The angular sensitivity in this direction is inversely proportional to chord width. An orthonormal Hadamard basis lets us give this residual extra weight during training while retaining supervision on the other seven directions. In five-fold image-level development experiments on 372 anteroposterior radiographs, we compared three objectives across 45 fits of the same landmark network. The three-seed corner ensemble reduced pooled symmetric percentage error from 18.45\% with coordinate supervision to 17.20\% with geometry weighting. Adding a soft pairwise-angle term yielded 16.89\%, a paired difference of $-1.56$ percentage points from coordinate supervision (95\% image-bootstrap interval $[-2.51,-0.61]$). The interval for the increment from adding the soft term included zero. In a single-fold comparison, direct unit-chord supervision also reduced error, and its difference from geometry weighting was uncertain. Across single-seed runs, rotational residual error fell by 36.1\% while normalized corner error rose from 2.60\% to 2.71\%. Perturbing one vertebra at a time also showed when a small rotation changes the selected end pair. These results identify a measurable trade-off between landmark fidelity and agreement with a specified geometric reader. The evaluation concerns an internal image-level split and landmark-derived target angles.
\end{abstract}

\section{Introduction}
\begin{figure}[htbp]
\centering\includegraphics[width=0.94\linewidth]{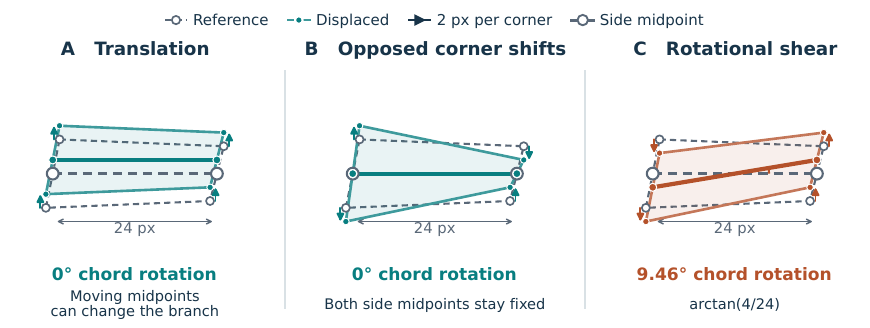}
\caption{Three equal-size errors on one schematic vertebra. Each corner moves two pixels from the dashed reference to the solid outline. The thick line joins the two side midpoints. Translation (A) can affect curve branching. The opposed shifts in (B) leave both endplate midpoints and chord unchanged. The active displacement (C) rotates this 24-pixel chord by $\arctan(4/24)=9.46^\circ$.}
\label{fig:modes}
\end{figure}

Consider two errors on the same vertebra. A network shifts all four corners two pixels upward, or it shifts the corners on opposite sides in opposite vertical directions. The corner displacement can be identical in magnitude. The first error translates the vertebra, whereas the second changes the tilt of the line joining its left and right sides. A model trained to predict the four points pays for both errors. A subsequent angle calculation responds quite differently to them (Fig.~\ref{fig:modes}). Its response also depends on whether the vertebra supplies an endpoint of a measured curve.

Vertebra-focused detectors estimate centers and corner offsets from frontal radiographs \cite{yi}. Other systems predict vertebral tilt fields \cite{tilt} or refine bending directions in a second network \cite{liang}. VLTENet supplies an especially close comparison: it derives a unit tilt vector from the same two side midpoints used here and supervises an output head for that vector \cite{vlte}. Guo et al. found that improving angle accuracy through mask--keypoint consistency could increase corner error \cite{guo}. Anisotropic landmark losses also have a precedent in face alignment, where ADNet penalizes the normal and tangential errors of individual points differently \cite{adnet}. Here the relevant displacement is shared by four points: a common translation can give all four corners a normal error while leaving the chord direction unchanged.

Other scoliosis models constrain anatomical distances and vertebral order \cite{yang}, quantify endplate wedging \cite{shi}, or regress a latent contour in place of four corners \cite{curvnet}. CVS-Spine combines a B-spline centerline with features from segmented vertebrae to estimate the angle \cite{cvsspine}. Our model retains the corner representation. We ask which joint corner displacements change the reading and whether a network can use that distinction during training.

We address this question for a discrete reader derived from a participant implementation associated with AASCE2019 \cite{challenge,reader}. We derive the coordinate direction that rotates each vertebral chord, then change the loss in a fixed network to test whether this distinction survives training. We measure the final hard three-angle output and the landmark errors paid for the improvement. A separate perturbation experiment connects a local chord rotation to a change in the selected end vertebrae. We compare the objectives under a common reader, show the predicted vertebrae and selected end pairs on radiographs, and place the results alongside recent curvature-estimation systems evaluated on other panels.

\section{From corner coordinates to a training objective}
\subsection{The direction seen by the angle reader}
Let $P_v\in\RR^{4\times2}$ contain the top-left, top-right, bottom-left and bottom-right corners of vertebra $v$, in that order. The reader takes 17 vertebrae in anatomical order. Define the orthonormal Hadamard vectors
\begin{equation}
\begin{split}
h_0&=(1,1,1,1)^\top/2,\qquad h_1=(1,1,-1,-1)^\top/2,\\
h_2&=(-1,1,-1,1)^\top/2,\qquad h_3=(1,-1,-1,1)^\top/2.
\end{split}
\end{equation}
The chord $d_v=P_v^\top h_2$ joins the left side midpoint to the right side midpoint. Set $w_v=\|d_v\|_2>0$, $a_v=d_v/w_v$ and $n_v=(-a_{vy},a_{vx})^\top$. The reader takes $\arccos(a_u^\top a_v)$ for each pair in native image coordinates, including angles above $90^\circ$. Its first output is the largest of the 136 unordered pair angles. A predicate on the sequence of 34 endplate midpoints and a height test on the selected vertebrae determine the branch for the two secondary angles. The reference output for each image is the value of this same function on its annotations. It measures agreement with a landmark-derived convention.

For a corner displacement $E$, the differential of the chord orientation $\phi_v$ is
\begin{equation}
 D\phi_v[E] = \frac{n_v^\top E^\top h_2}{w_v}
             = \frac{\langle E,h_2 n_v^\top\rangle_F}{w_v}.             \label{eq:deriv}
\end{equation}
Indeed, moving by $t h_2 n_v^\top$ changes the chord to $w_va_v+t n_v$, so that its rotation is exactly $\arctan(t/w_v)$. Three orthogonal directions, $h_3 e_x^\top$, $h_3 e_y^\top$ and $h_2a_v^\top$, preserve both endplate midpoints and chord orientation as long as the chord retains positive length along $a_v$. The remaining four directions, $h_j e_k^\top$ for $j\in\{0,1\}$ and $k\in\{x,y\}$, preserve the entire pair-angle matrix but can change the branch through the midpoint positions. These seven directions and $h_2 n_v^\top$ form an orthonormal basis under the Frobenius inner product. Thus one coordinate determines the local tilt response. Several vertebrae, their relative rotations and the discrete branch determine the final curvature measurement.

\subsection{A shared landmark network}
\begin{figure}[htbp]
  \centering\includegraphics[width=\linewidth]{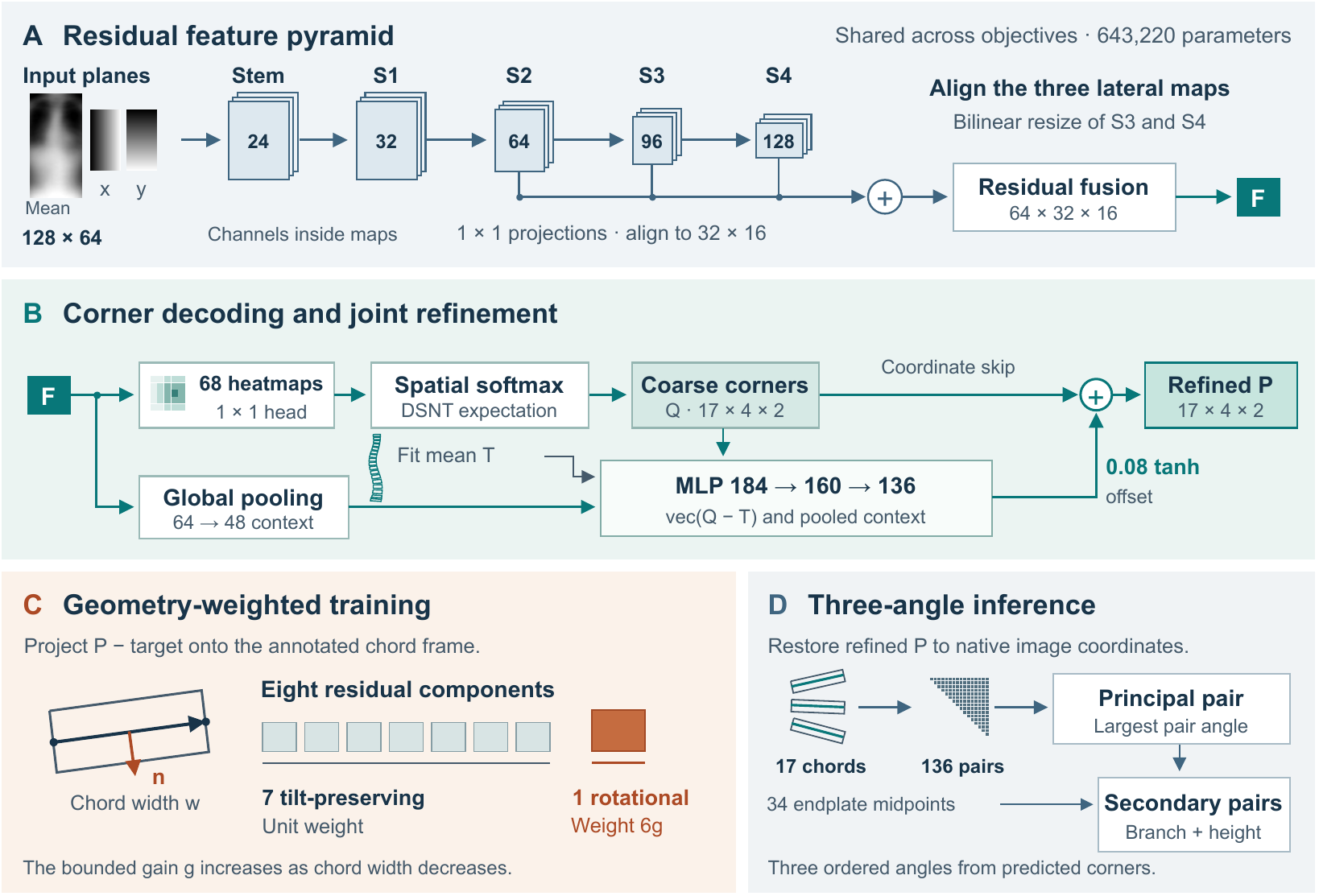}
  \caption{One predictor serves all three training objectives. (A) The network appends horizontal and vertical coordinate planes to each radiograph. The grayscale inset shows the mean of the 329 standardized fitting images in fold 0. (B) The fused map $F$ supplies both coarse corners $Q$ and image context. The inset beside $T$ shows the mean corner template from the same fitting fold. (C) Geometry-weighted supervision uses the annotated chord frame to emphasize rotational error. (D) At inference, the hard reader uses the refined corners at native image scale. The heatmap and chord drawings are schematic.}
  \label{fig:arch}
\end{figure}

The network takes a standardized $128\times64$ grayscale image and two coordinate planes \cite{coordconv}. A residual encoder produces four feature scales. The last three are projected to 64 channels, resized to $32\times16$ and added. The fused map produces 68 spatial heatmaps. A differentiable spatial-to-numerical transform (DSNT) \cite{dsnt} takes the expectation of each corner location over pixel centers. A small residual MLP uses the coarse coordinates relative to the fitting-fold mean shape and a pooled image feature to refine all 68 corners jointly (Fig.~\ref{fig:arch}). All three objectives use this 643,220-parameter network and the same forward pass. The predicted corners can be retained for inspection after inference.

\subsection{Three training objectives}
For training we express the normalized corner residual $E_v=\wh P_v-P_v^*$ in a canonical frame of width 256 and height 512. Let $\xi_{vj}$ denote its coefficient in the reference basis, ordered with the rotational component last. The frame and width come from the original fitting annotation. The residual uses the augmented target. Let $\rho_\beta$ be Smooth-$L_1$ with transition $\beta$ ($x^2/(2\beta)$ inside the transition, $|x|-\beta/2$ outside). Each arm includes the same center term $C$, one quarter of the mean componentwise $\rho_2$ between coarse DSNT centers and annotated four-corner means.
The coordinate control minimizes $C$ plus the componentwise mean of $\rho_2(E_v)$. In the Hadamard-weighted arm the corner term becomes
\begin{equation}
 L_H = \frac{1}{17}\sum_{v=1}^{17}\frac{1}{8}\!
 \left[\sum_{j=1}^7\rho_2(\xi_{vj})+
 6 g_v\rho_2(\xi_{v8})\right],\qquad
 g_v=\clip\!\left(\frac{\bar w_{\rm fit}}{\max(w_v,6)},0.5,3\right).  \label{eq:weight}
\end{equation}
Here $\bar w_{\rm fit}$ is the mean width across the fold's fitting vertebrae. The weighting increases the cost of rotational error at narrow levels. Equation~\eqref{eq:deriv} motivates the direction and the dependence on width. A squared small-angle loss would scale as $1/w_v^2$. The bounded $1/w_v$ gain is a milder design choice. The other directions retain a cost, including those that affect the secondary-curve branch. Coordinatewise Smooth-$L_1$ also changes under a rotation of its basis outside the quadratic region. This experiment tests the change of basis, the factor six and the bounded inverse-width factor together.

The third arm adds a soft angle term to $C+L_H$. Let $\Theta(P)$ collect the 136 upper-triangular pair angles in degrees in the canonical frame, with dot products clipped to $[-0.99999,0.99999]$. Define $m(\Theta)=2\log\sum_j\exp(\Theta_j/2)$ in degrees. For epoch $e$, the additional loss is
\begin{equation}
 \clip((e-12)/18,0,1)\left[
 0.10\,\avg_j\rho_1(\wh\Theta_j-\Theta_j^*)+
 0.05\,\rho_1\big(m(\wh\Theta)-m(\Theta^*)\big)\right].
\end{equation}
Both terms compare predictions to the corresponding reference geometry. The soft maximum approximates the largest pair angle and gives weight to its close competitors. The discrete decisions for the two secondary curves enter through evaluation.

\section{Experiments}
\subsection{Radiographs, partitions and endpoints}
We used the 481-image training partition of the AASCE spinal radiograph collection \cite{challenge,shape}, which provides four corner landmarks for each vertebra from T1 through L5. To preserve untouched holdout sets, 77 images within this release were excluded from this cross-validation study, and neither the 128-image public evaluation set nor the 98-image challenge test set was accessed. Thirty-two images previously used to verify the angle-reader implementation against reference annotations were kept strictly on the fitting side of every fold so that they never entered out-of-fold evaluation. The remaining 372 images were assigned once each to five reporting folds of sizes 75, 75, 74, 74, and 74, yielding fitting sets of 329 or 330 images per fold. Because all evaluations are conducted via cross-validation within the public training partition, the resulting metrics characterize internal development performance rather than blinded external generalization.

We checked the fold membership against the numeric record field present in one family of source filenames. Of 372 reporting images, 357 had a parseable field. For 252, at least one fitting image in the same fold bore the same field. These counts are evidence of record overlap under this key. The data holder has not verified patient identities. The comparison therefore estimates a paired difference on this image-level split. Shared subjects or examinations may narrow an image-bootstrap interval and inflate apparent generalization.

For radiograph $i$, define $u_i=\sum_{k=1}^3 |\hat y_{ik}-y_{ik}|$ and $v_i=\sum_{k=1}^3 (|\hat y_{ik}|+|y_{ik}|)$, with $y=g(P^*)$ and $\hat y=g(\wh P)$. The neural loss experiment uses pooled symmetric mean absolute percentage error,
\begin{equation}
 S_{\rm pool}=100\frac{\sum_i u_i}{\sum_i v_i},\qquad
 D_{\rm image}=\frac{100}{N}\sum_i\frac{u_i}{v_i}. \label{eq:smape}
\end{equation}
The second statistic, mean per-image SMAPE, is used for perturbation summaries and the higher-resolution comparisons below. The factor is 100 in both cases. $S_{\rm pool}$ gives more weight to images with a larger combined angle magnitude. We also report three-angle mean absolute error (MAE) in degrees, and the mean Euclidean corner error divided by the native image diagonal (NME). These measures let us distinguish a better angle from a closer corner.

\subsection{Matched fits and perturbation tests}
For each of five folds and three seeds, all three objectives shared the same initial parameters, mini-batch order, augmentation draws and optimizer schedule. Each model trained from scratch for 60 epochs. We scored the final epoch and formed an ensemble by averaging the three seeds' out-of-fold corner coordinates before applying the nonlinear reader. Nominal 95\% percentile intervals for ensemble contrasts used 10,000 paired image resamples, recomputing the pooled ratio within each draw. They condition on the fitted models. We report the standard deviation of the three individual seed scores separately.

To compare geometry weighting with direct direction supervision, we trained four additional models on CPU using fold~0 (329 fitting and 75 reporting images), a fixed random seed (Seed~1), and the same 60-epoch schedule. The fourth objective adds $\frac{1}{17}\sum_v\|\hat a_v-a_v^*\|_1$ to the coordinate control, with a fixed coefficient of one. Both unit vectors come from the predicted or annotated corners in the canonical pixel frame. VLTENet supervises a separate tilt-vector head \cite{vlte}. Deriving the vector from our corners lets us compare the two forms of supervision without changing the predictor. These fits supply the single-fold comparison and the model used for the radiograph illustrations.

Two perturbation tests used the 372 reference landmark sets. In the first, every vertebra received an independently sampled direction in a specified subspace. The Frobenius norm per vertebra was $2\sigma$, hence every configuration at a given $\sigma$ moved the corners by the same two-dimensional RMS displacement. We evaluated 20 draws per radiograph at each of five displacements ($0.5,1,2,4,8$ native pixels). In the second test, we changed just one vertebra, in each direction along $h_2n_v^\top$, with $t=\pm2\sigma$ for $\sigma=1,2,4$ native pixels. This gives 34 signed changes per radiograph and scale, with all other vertebrae held fixed.

A single-level perturbation permits a direct check on the end-pair selection. Let $m$ be the gap between the largest and second-largest unperturbed angles among distinct unordered pairs. Rotating one chord through $\delta=|\arctan(2\sigma/w_v)|$ can change any pair angle by at most $\delta$ in angular distance. The leading pair can lose at most $\delta$ while another gains at most $\delta$. Thus
\begin{equation}
                     m>2\delta \quad\Longrightarrow\quad
           \text{the principal end pair is unchanged}.          \label{eq:margin}
\end{equation}
We used degrees on both sides, required an extra margin of $10^{-9}$ degrees to exceed the implementation's tie tolerance, and reread the perturbed corners. This is a sufficient condition for the principal pair. Its angle value itself can still change.

\section{Results}
\subsection{Equal pixel displacements have different effects}
At a 4-pixel RMS shift of all 17 vertebrae, active rotational shear produced a mean per-image error of 10.498\% and a three-angle MAE of $4.345^\circ$. The seven directions orthogonal to rotation produced 0.332\% and $0.127^\circ$ at the same displacement. Full-space isotropic perturbations gave 4.584\% and $1.781^\circ$. Motion confined to the three exact invariant modes left all three readings unchanged to within $1.05\times10^{-11}$ degrees across the tested scales. The branch-sensitive modes account for the small, nonzero mean error in the seven-dimensional complement. Within individual radiographs, mean angular sensitivity per pixel at T1--T3 was $1.48\pm0.20$ times that at L3--L5.

\begin{figure}[htbp]
\centering\includegraphics[width=\linewidth]{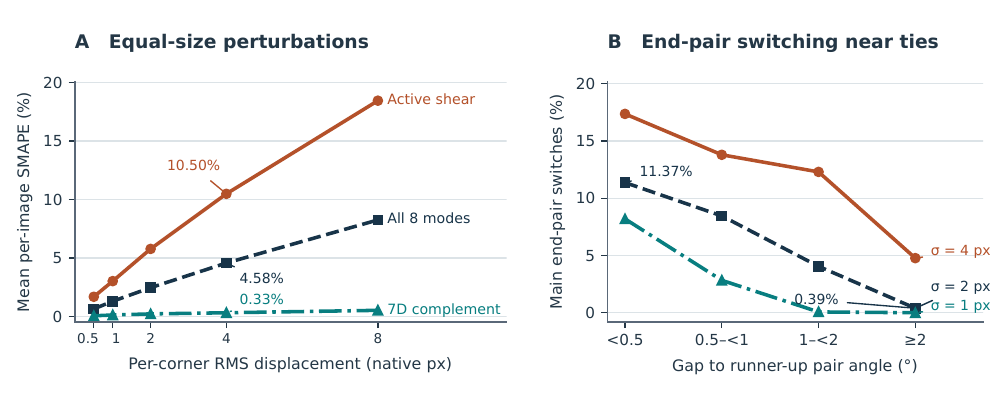}
\caption{How a corner displacement reaches the hard reader. (A) All levels are perturbed at the same RMS displacement. Each point averages 20 draws on each of 372 radiographs. (B) One level at a time is tilted in each direction. The main end-pair switch rate is shown against the gap to the next largest pair angle. The four bins contain 127, 81, 82 and 82 images, respectively. Each image contributes 34 paired-direction trials per scale.}
\label{fig:geometry}
\end{figure}

In the single-level test, a 2-pixel shift switched the principal pair in 491 of 4,318 trials (11.37\%) when the gap was below $0.5^\circ$, and in 11 of 2,788 trials (0.39\%) when it was at least $2^\circ$ (Fig.~\ref{fig:geometry}B). Mean absolute change in the principal \emph{angle} was similar in those two bins ($0.192^\circ$ and $0.182^\circ$). End-pair identity is therefore more fragile than the angle value near a tie. Across all 37,944 perturbations, 5,272 satisfied Eq.~\eqref{eq:margin} and all 5,272 retained their original pair. Among the other 32,672 perturbations, 2,888 switched. These are repeated interventions on known reference shapes, with 372 distinct radiographs.

\subsection{Geometry weighting trades coordinate precision for angle accuracy}
Across the three seeds, the weighted objective reduced pooled angle error from 20.02\% to 17.53\%, and weighting plus soft angles reduced it to 17.17\% (Table~\ref{tab:main}). The corresponding ensemble scores were 18.45\%, 17.20\% and 16.89\%. Weighting improved ensemble error by 1.25 percentage points relative to coordinate supervision (95\% interval for weighted minus coordinate $[-2.18,-0.31]$). With soft angles included the difference was $-1.56$ points $[-2.51,-0.61]$, and ensemble MAE fell from $9.26^\circ$ to $8.25^\circ$ (paired difference $-1.00^\circ$, $[-1.45,-0.55]$). The incremental difference for adding soft angles to weighting was $-0.312$ points $[-0.634,+0.041]$. The three arms did not include a soft-only objective.

\begin{table}[htbp]\centering
\caption{Matched loss comparison on 372 out-of-fold radiographs. Each arm used five folds and three seeds. The single-model columns summarize seed-level results. RMS residuals are computed per image in canonical pixels and then averaged. The ensemble averages corner predictions before calculating angles.}
\label{tab:main}\small
\begin{tabular}{@{}lrrrrrr@{}}\toprule
 & \multicolumn{4}{c}{Single model, mean across three seeds} & \multicolumn{2}{c}{Ensemble}\\
 \cmidrule(lr){2-5}\cmidrule(l){6-7}
Objective & $S_{\rm pool}$ (\%) & MAE ($^\circ$) & Active RMS & NME (\%) & $S_{\rm pool}$ (\%) & MAE ($^\circ$)\\\midrule
Coordinate & $20.02\!\pm\!0.25$ & 10.58 & 5.77 & 2.60 & 18.45 & 9.26\\
Weighted & $17.53\!\pm\!0.15$ & 8.64 & 3.72 & 2.69 & 17.20 & 8.29\\
Weighted + soft & $17.17\!\pm\!0.15$ & 8.58 & 3.69 & 2.71 & 16.89 & 8.25\\\bottomrule
\end{tabular}
\end{table}

\begin{figure}[htbp]
\centering\includegraphics[width=\linewidth]{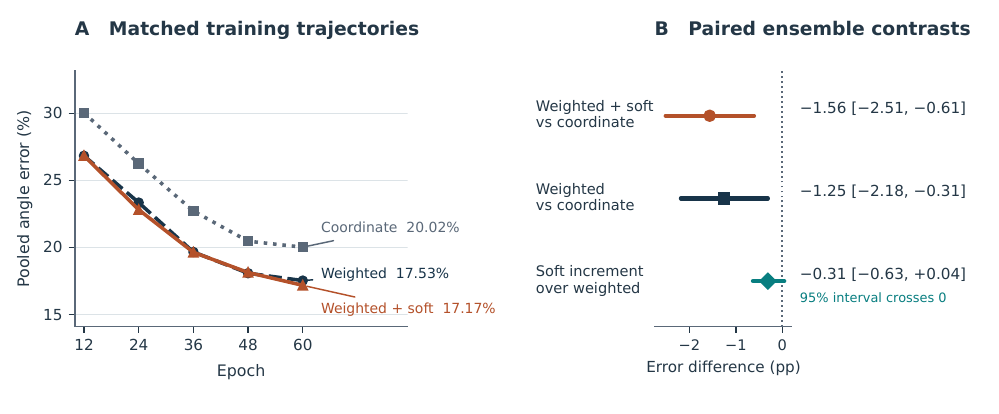}
\caption{Angle error during training and after ensembling. (A) Each curve averages three seed-specific pooled out-of-fold scores. Epoch 60 supplies the reported predictions. (B) A point is the paired difference between two three-seed corner ensembles. Horizontal lines show 95\% image-bootstrap intervals. Negative values favor the first objective named.}
\label{fig:learning}
\end{figure}

The gain had a measurable cost. For individual models, RMS error along the active rotational direction fell from 5.77 to 3.69 canonical pixels (36.1\%), while RMS error in the other seven directions rose from 14.56 to 15.40. Mean corner NME rose by 0.11 percentage points. The component of displacement amplified by the reader improved even as overall localization became less accurate. Figure~\ref{fig:learning} shows the training trajectories and the uncertainty in each final ensemble contrast.

\subsection{Does a direct tilt loss explain the gain?}
On the 75 reporting images from fold~0, direct supervision of the unit chord reduced pooled error from 23.30\% to 19.84\% (Table~\ref{tab:direction}). Geometry weighting alone gave the lowest angle error, 18.51\%, while the direct loss gave the lowest corner error, 2.79\% of the image diagonal. Adding soft angles to geometry weighting increased pooled error on this fold to 19.14\%. In 10,000 paired image resamples the difference between weighting alone and direct tilt supervision was $-1.33$ percentage points (95\% interval $[-4.22,1.47]$). The direction control accounts for much of the gain over ordinary coordinate supervision. This single fold leaves the ordering of the two direction-sensitive losses uncertain.

\begin{table}[htbp]\centering
\caption{Single-fold comparison on the same 75 radiographs. Each row is a fold-0 model trained on CPU with the same seed, initialization, augmentation, and 60-epoch schedule. The two percentage errors use Eq.~\eqref{eq:smape}.}
\label{tab:direction}\small
\begin{tabular}{@{}lrrrr@{}}\toprule
Training objective & $S_{\rm pool}$ (\%) & $D_{\rm image}$ (\%) & MAE ($^\circ$) & NME (\%)\\\midrule
Coordinate & 23.30 & 23.87 & 12.34 & 2.83\\
Coordinate + direct unit chord & 19.84 & 21.33 & 10.32 & 2.79\\
Geometry-weighted & 18.51 & 20.00 & 9.04 & 2.83\\
Geometry-weighted + soft angles & 19.14 & 20.37 & 9.46 & 2.84\\\bottomrule
\end{tabular}
\end{table}

\subsection{Published curvature-estimation systems}
Table~\ref{tab:frontier} separates the internal loss comparison from published benchmark and clinical results. Among systems evaluated on the public 128-image AASCE panel, LaNet uses a bending-segment rule to turn landmarks into angles, RouterNet routes predictions from a root landmark to individual corners, and S-Former includes a separate spine-region crop. Architectural performance on that panel answers a different question from the change in loss on our 372 development images. The reader itself matters. On the same validation panel, LaNet reports 6.81\% SMAPE with a conventional angle calculation and 4.51\% with its bending-segment rule (Table~VII in Ref.~\cite{lanet}).

\begin{table}[htbp]\centering
\caption{Curvature-estimation results by evaluation protocol. The first row gives the three-seed corner ensemble in this study. The remaining rows are the authors' reported values. CMAE denotes circular mean absolute error where specified by the authors. A pooled ratio, a mean per-image percentage, and an error in degrees summarize different quantities.}
\label{tab:frontier}\small
\begin{tabularx}{\linewidth}{@{}>{\raggedright\arraybackslash}p{3.60cm}>{\raggedright\arraybackslash}p{4.05cm}X@{}}\toprule
Method & Evaluation panel & Target and reported result\\\midrule
Weighted + soft (ours) & 372 internal out-of-fold images & Three reader angles: $S_{\rm pool}$ 16.89\%, MAE $8.25^\circ$\\\midrule
LaNet (2024) \cite{lanet} & AASCE public 128 & Three angles: SMAPE 4.51\%, CMAE $2.06^\circ$\\
VLTENet (2023) \cite{vlte,lanet} & AASCE public 128 & Three angles: SMAPE 5.44\%, CMAE $2.51^\circ$\\
S-Former (2025) \cite{sformer} & AASCE public 128 & Three angles: SMAPE 7.35\%, CMAE $3.50^\circ$\\
RouterNet (2026) \cite{routernet} & AASCE public 128 & Three angles: SMAPE 6.24\%, CMAE $2.69^\circ$\\
\addlinespace
CurvNet (2026) \cite{curvnet} & AASCE public 128 & Maximum angle only: SMAPE 5.05\%\\
VCRLD-Net (2026) \cite{vcrld} & Curated AASCE series & SMAPE 5.01\%, CMAE $2.38^\circ$\\
SpineNET (2026) \cite{spinenet} & 150 external radiographs & Surgeon consensus: Cobb MAE $1.44^\circ$\\\bottomrule
\end{tabularx}
\end{table}

The LaNet and VLTENet entries share the public validation panel identified in Table~II of Ref.~\cite{lanet}. S-Former and RouterNet report the 128-image results in their Tables~2 and~4, respectively. CurvNet also gives separate SMAPEs of 12.48\%, 6.27\% and 14.04\% for the proximal thoracic, main thoracic and thoracolumbar/lumbar angles, respectively (Table~3 in Ref.~\cite{curvnet}). VCRLD-Net uses a curated AASCE series of 554 images in total. The clinical SpineNET study used another center with whole-spine, thoracic and lumbar views. Its MAE measures agreement with three surgeons. Our five-fold comparison isolates loss-induced differences under one landmark-derived convention on the internal development split; the public 128-image and hidden 98-image challenge partitions were intentionally left untouched, so Table~\ref{tab:frontier} provides protocol context rather than a direct benchmark ranking.

The same 372 reporting images and native reader also allow a direct comparison in degrees with two higher-resolution landmark systems evaluated on the same partition. The mean individual-model three-angle MAE over three seeds was $4.41^\circ$ for a hierarchical landmark regressor with a ConvNeXt-S encoder and $4.52^\circ$ for a Yi-style center-offset detector \cite{yi}, versus $8.58^\circ$ for our compact network with geometry weighting and soft angles. The larger systems used ImageNet pretraining, larger inputs and 180 training epochs. They establish the accuracy gap for the complete models on the same images, while Table~\ref{tab:main} isolates the loss change.

On a separate 481-image development panel we evaluated an adaptation of RouterNet alongside these two families of landmark systems. Using the same fold roles and native reader, image-mean SMAPE was 14.20\% for RouterNet, 15.91\% for our hierarchical landmark regressor and 11.35\% for the vertebra-focused detector \cite{yi}. The corresponding three-angle MAEs were $6.71^\circ$, $6.88^\circ$ and $5.06^\circ$. All three systems produced readings for all 481 radiographs. Training budgets and checkpoint selection differed across systems. The geometry-weighted objective was tested on the separate 372-image development protocol.

\subsection{Reading the model's output on real radiographs}
\begin{figure}[htbp]
\centering\includegraphics[width=0.90\linewidth]{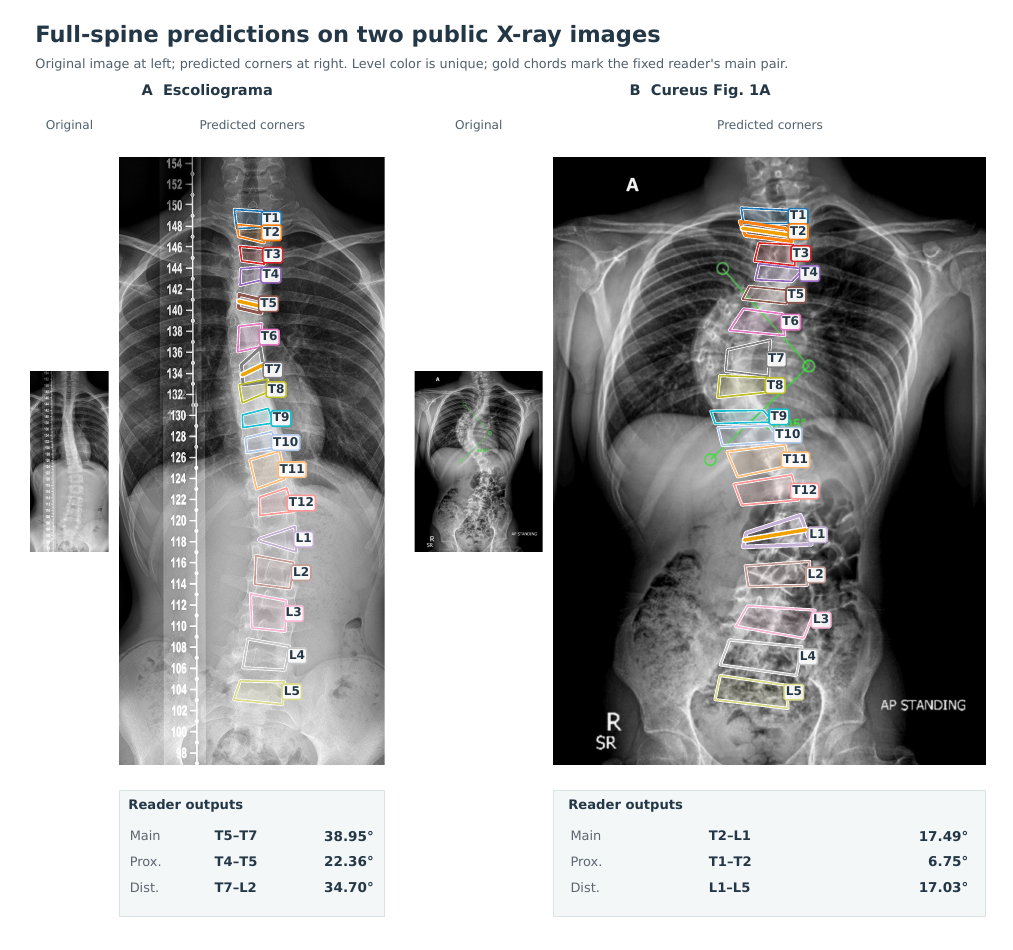}
\caption{The same fold-0 geometry-weighted model with the soft angle term applied to two illustrative public frontal radiographs outside quantitative evaluation. Each transparent quadrilateral connects four predicted corners (rather than representing a pixel-wise mask). The model assigns level labels in T1--L5 order, and the hard reader selects the end pairs. First image adapted from \href{https://commons.wikimedia.org/wiki/File:Escoliograma.jpg}{\emph{Escoliograma.jpg}}, Departament de radiologia (2011), classified as public domain by Wikimedia Commons. Second image adapted from Fig.~1A of Baloch et al.\ \cite{baloch} under \href{https://creativecommons.org/licenses/by/4.0/}{CC BY 4.0}. Both images were cropped and overlaid. The green measurement ($88.05^\circ$) on the second film is the source authors' annotation.}
\label{fig:cases}
\end{figure}

Figure~\ref{fig:cases} follows the predictions from the image to the angle on two publicly available radiographs chosen for visibility of the thoracic and lumbar spine. We downsampled a crop of each frontal view to $128\times64$ for inference and mapped the 68 predicted corners back onto the radiograph. The second figure panel retains 80 extra pixels at the right margin of the source image. Joining the four corners at each level gives a quadrilateral that lets the reader inspect both position and tilt. The shaded region only interpolates between predicted points. The labels T1--L5 denote the ordered model outputs. Their positions can be checked against the underlying anatomy.

On the first film the largest angle between predicted chords was $38.95^\circ$, between levels labeled T5 and T7 by the model. On the second film the predicted chain misses the lateral excursion of the thoracic curve. The model reads $17.49^\circ$ between T2 and L1, while the source authors marked an $88.05^\circ$ Cobb angle. Their endplate lines were already visible in the input. The source measures endplate tilt and our reader measures the angle between vertebral chords. The missed bend is also visible directly on the film. The three public case images supply no independent set of 68 reference corners and are outside the quantitative evaluation.

\begin{figure}[htbp]
\centering\includegraphics[width=0.92\linewidth]{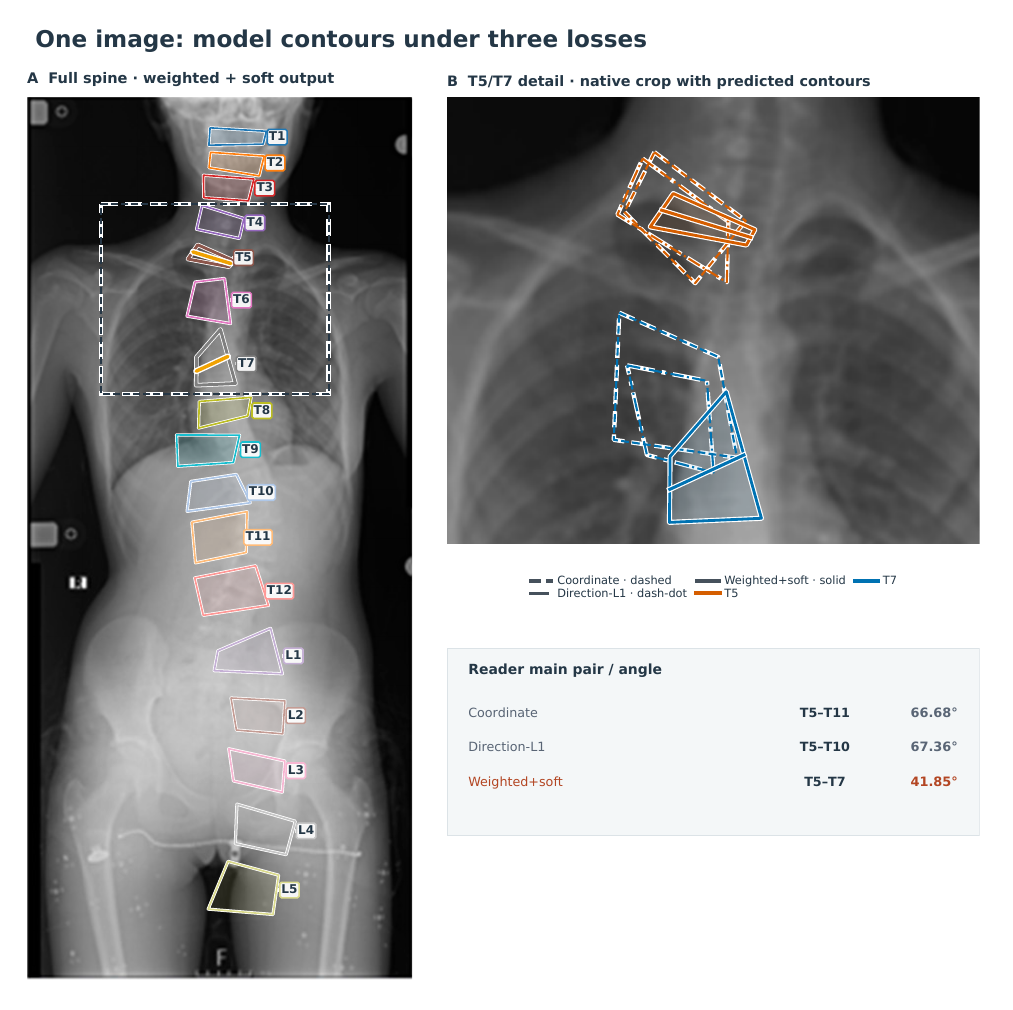}
\caption{Local geometry on a low-contrast standing radiograph under three fold-0 models trained with different objectives. (A) Predicted vertebral quadrilaterals from the geometry-weighted model with the soft angle term, mapped back to native image coordinates; gold chords mark the reader-selected main pair (T5--T7). (B) Enlarged native crop around T5 and T7 comparing the three training objectives by line style, together with each model's fixed-reader main pair and angle. The low-contrast radiograph is shown without enhancement and has no independent 68-corner reference. Adapted from Fig.~4A of Aktan et al.\ \cite{aktan}, distributed under \href{https://creativecommons.org/licenses/by/4.0/}{CC BY 4.0}.}
\label{fig:case-detail}
\end{figure}

Figure~\ref{fig:case-detail} uses a third film to show how a reader can trace a predicted angle back to its four-corner geometry. The coordinate model selected T5 and T11 ($66.68^\circ$), the direct-direction model selected T5 and T10 ($67.36^\circ$), and the geometry-weighted model with the soft term selected T5 and T7 ($41.85^\circ$). Changing the training objective moved the lower endpoint on the same film. The source reports a dominant thoracolumbar/lumbar curve based on a multi-view clinical assessment \cite{aktan}. The overlay also makes a wide-field level-assignment shift immediately visible: the quadrilateral labeled T1 lies in the cervical region and L5 lies below the pelvis, illustrating how retaining explicit corner predictions exposes field-of-view misalignment that a scalar angle output alone would conceal.

\begin{figure}[htbp]
\centering\includegraphics[width=\linewidth]{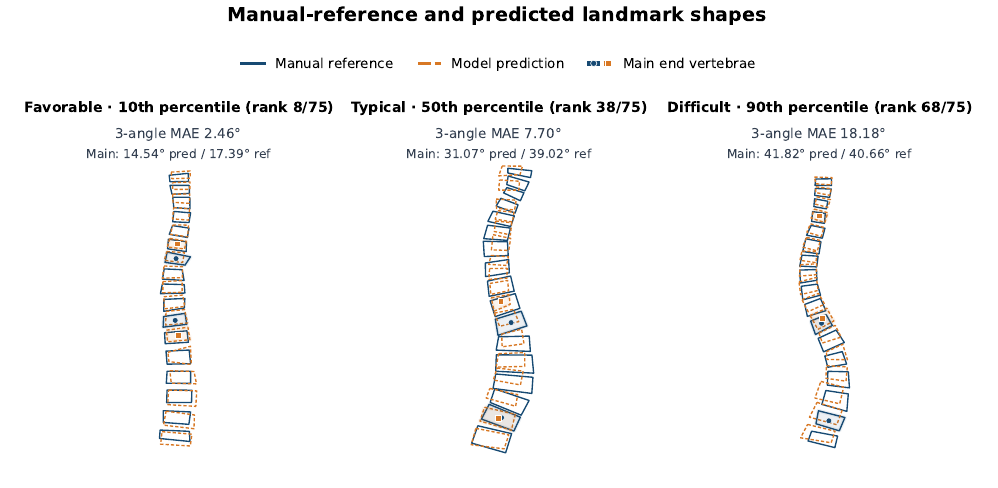}
\caption{Predicted and manually annotated vertebral quadrilaterals for three reporting images from fold~0. The ranks are the nearest ranks for the 10th, 50th and 90th percentiles of three-angle MAE for the geometry-weighted model with the soft term. Both sets of landmarks are shown in the same coordinate frame for each image. Only landmark geometry is displayed.}
\label{fig:held}
\end{figure}

The public case images show where the predictions fall on the film. We also compared predicted corners with manual reference corners in the 75-image reporting fold (Fig.~\ref{fig:held}). Sorting by three-angle MAE gave errors of $2.46^\circ$, $7.70^\circ$ and $18.18^\circ$ at ranks 8, 38 and 68. In the last case the principal angle was within $1.16^\circ$ of the reference, but errors of $22.96^\circ$ and $30.42^\circ$ on the two secondary angles determined the overall result. Inspecting only the largest angle would conceal the error on this film.

\FloatBarrier
\section{Discussion}
A corner can be anatomically misplaced and still supply the correct chord angle. The perturbation and learning results describe two different aspects of that observation. Perturbing reference corners shows exactly which directions the specified reader amplifies and when a branch or pair can change. Changing supervision on a fixed network shows that the regressor can trade errors among those directions. In this experiment the trade improved agreement with the reader while sacrificing a small amount of point accuracy. A downstream application that requires precise vertebral contours would need both outcomes.

The local derivative also explains a limit of loss design. The same pixel displacement rotates a narrow vertebra more than a wide one, but this rotation has no effect on the principal reading unless it changes or belongs to the leading pair. Near a tie, two neighboring vertebrae can exchange roles with little effect on angle magnitude. Weighting shear by inverse width uses the first fact without requiring a hard choice of end vertebra at training time. The secondary readings still depend on midpoint geometry. Penalizing only the single rotational mode would discard that information.

The study isolates a training \emph{recipe} in one compact network. There was no equal-weight Hadamard control or constant-gain rotational control to identify which ingredient of Eq.~\eqref{eq:weight} was necessary. The incremental soft-angle interval contains zero, and the soft training angles are computed in a common aspect ratio, while the hard evaluator restores each image's native dimensions. The fixed low resolution makes loss contrasts inexpensive to match, but discards edge detail available to a higher-resolution detector (Appendix~\ref{sec:context}). Whether the same rotational weighting transfers directly to a higher-resolution detector remains to be tested in a matched loss ablation.

The target was obtained by running the same reader on predicted and annotated corners. It contains no independent clinician choice of end vertebrae. The source reader, the aggregation rule and the image-level split are therefore part of the definition of the result. Repeated record keys across fitting and reporting images constrain absolute performance estimates and the conditional intervals. The most informative test of transfer would retain a higher-resolution detector's training and decoding recipe, change only the loss, and evaluate on independently grouped radiographs with prespecified landmark and angle endpoints.

\section{Conclusion}
The four corners of one vertebra contain one local direction that rotates its left-to-right chord and seven that leave its continuous orientation unchanged. A loss that emphasizes this rotational component shifted a landmark network's error toward the less angle-sensitive directions and lowered three-angle error on the development panel. In this setting the direction of a localization error matters enough to reverse the ranking suggested by average corner accuracy.

\paragraph{Data and ethics.}
Training images and annotations come from the publicly released AASCE/SpineWeb Dataset~16 collection \cite{challenge,shape}. The illustrative radiographs are reproduced under their stated open licenses and credited in Figs.~\ref{fig:cases} and~\ref{fig:case-detail}. This study analyzed existing retrospective and publicly available images and involved no new participant recruitment, clinical intervention, or human reading sessions; accordingly, no new institutional ethics approval was required. Full model, optimization, and reader implementation specifications are provided in Appendix~\ref{sec:impl}.

\paragraph{Competing interests.}
The author declares no competing financial or non-financial interests.

\paragraph{Acknowledgments.}
Artificial intelligence tools were used solely to assist with English language polishing, research and ideation, identifying related work, and accelerating code development and visualization. All technical content, code, visualizations, analyses, and conclusions were thoroughly checked and verified by the author, who assumes full responsibility for the manuscript.

\appendix
\section{Implementation details}\label{sec:impl}
\paragraph{Network and optimization.}
After per-image standardization, normalized horizontal and vertical coordinate channels are concatenated with the grayscale input. The stem uses a $3\times3$ convolution with stride 2 and 24 output channels (followed by batch normalization and GELU). The four residual stages have widths 32, 64, 96 and 128. Lateral projections from the last three stages have width 64. After fusion, a $1\times1$ convolution yields one logit plane per corner. The spatial softmax uses a multiplier of 2.5 and evaluates positions at normalized cell centers. The context branch pools the fused feature map and maps 64 features to 48. The refinement MLP maps the concatenated 136 relative coordinates and 48 context values through 160 hidden units to 136 residuals. It adds $0.08\tanh(\cdot)$ to the coarse normalized coordinates. The center term supervises the four-corner mean of the coarse predictions.

The 45 neural fits used PyTorch on the Apple Metal backend. We used AdamW with initial learning rate $2.5\times10^{-3}$, weight decay $10^{-4}$, batch size 64 (including the final short batch) and gradient norm limit 3. The cosine schedule ended at 5\% of the initial rate. Augmentation translated a batch by up to three horizontal and five vertical input pixels with replicated image borders, then applied per-image gain and bias jitter. Augmented normalized targets were clipped to $[0.005,0.995]$. Reference bases and widths were calculated before augmentation. Three fixed random seeds (Seeds~1--3) were shared across loss arms, using identical fold-specific initialization and deterministic data-order and augmentation generators. Each image had one out-of-fold prediction per seed and arm. The final evaluation used the three angles in reader order. Single-seed pooled errors across Seeds~1--3 were $(20.234,20.075,19.746)$, $(17.695,17.473,17.413)$, and $(17.336,17.034,17.146)$ percent for coordinate, weighted, and weighted-plus-soft supervision, respectively.

\paragraph{Reader implementation.}
Our implementation uses double-precision geometry adapted from the deposited participant reader \cite{reader}. Each native-pixel chord connects the average of its two left corners to the average of its two right corners. Angles use a clipped cosine with no acute-angle folding. Ties within $10^{-10}$ degrees are resolved by the first maximum in column-major order of the full pair matrix. The S predicate compares 34 endplate midpoints to the line between the first and last midpoint. With no S configuration, secondary angles use the first and last vertebrae as boundary references. In an S configuration, the vertical positions of the main pair determine which prefix or suffix search supplies each secondary angle. To avoid division by zero during neural evaluation, any predicted chord of length at most $10^{-6}$ native pixels has its normalized right-side horizontal coordinates offset by $10^{-3}$ prior to angle computation.

\paragraph{Perturbation and overlap counts.}
At each native-pixel displacement in the all-level test, Gaussian directions within each subspace were normalized independently for each vertebra. The axial component was reflected if a large negative displacement would threaten chord orientation in the invariant subspaces. The single-level experiment used both signs deterministically. It counted a switch whenever the unordered identities of the two principal end vertebrae changed. For nominal displacements of 1, 2 and 4 pixels, respectively, 434, 848 and 1,606 of 12,648 trials switched. Gap bins were descriptive. The separate filename audit compared the numeric record field to the set of fitting keys for each image's own fold. In fold order, 42, 53, 49, 55 and 53 reporting images matched fitting keys, out of 72, 70, 73, 72 and 70 parseable names. Identifiers and images are omitted from the aggregates.

\section{Surrogate-to-angle mismatch at higher resolution}\label{sec:context}
To examine whether surrogate landmark loss also disagrees with hard Cobb angle ranking in a higher-resolution detector, we analyzed a vertebra-focused (VF) detector \cite{yi} trained at $1024\times512$ across all 481 AASCE training radiographs. This pipeline sorts predicted vertebral centers vertically and uses a midpoint-based shape rule, whereas the controlled 372-image study preserves anatomical corner order and uses 34 endplate midpoints; both the reader convention and the aggregation in Eq.~\eqref{eq:smape} therefore differ across the two protocols. Across five folds and three random seeds (15 matched trajectories), we compared the checkpoint minimizing validation surrogate loss against the terminal epoch-50 checkpoint. In all 15 trajectories, the terminal epoch-50 checkpoint achieved lower out-of-fold mean per-image SMAPE, reducing mean $D_{\rm image}$ across seeds from 11.317\% to 9.467\%. Across the three seeds individually, the paired reductions were $-2.168$, $-1.283$, and $-2.099$ percentage points, with nominal 95\% bootstrap intervals resampling 221 filename-derived record groups of $[-2.854,-1.489]$, $[-1.925,-0.640]$, and $[-2.893,-1.339]$. The same ordering disagreement persisted across all 30 inner-validation combinations of seed, fold, and input pipeline. These observations show that minimizing an isotropic surrogate landmark loss does not guarantee optimal hard-reader angle accuracy even at high resolution, while leaving a matched transfer test of Eq.~\eqref{eq:weight} in higher-resolution detectors for future work.

\end{document}